\documentclass[10pt,conference]{IEEEtran}

\usepackage{amsmath}
\usepackage{amssymb}
\usepackage{array}
\usepackage{booktabs}
\usepackage{caption}
\usepackage{color}
\usepackage{comment}
\usepackage{float}
\usepackage{flushend}
\usepackage[toc, acronym]{glossaries} 
\usepackage{graphicx}
\usepackage{ifpdf}
\usepackage{keyval}
\usepackage{listings}
\usepackage{microtype}
\usepackage{moresize}
\usepackage{multirow}
\usepackage{paralist}
\usepackage{rotating}
\usepackage{setspace}
\usepackage{soul}
\usepackage{subfig}
\usepackage{tabularx}
\usepackage{url}
\usepackage{wrapfig}
\usepackage{xcolor}
\usepackage{xspace}
\usepackage{setspace}

\definecolor{listinggray}{gray}{0.95}
\definecolor{darkgray}{gray}{0.7}
\definecolor{commentgreen}{rgb}{0, 0.4, 0}
\definecolor{darkblue}{rgb}{0, 0, 0.4}
\definecolor{middleblue}{rgb}{0, 0, 0.7}
\definecolor{darkred}{rgb}{0.4, 0, 0}
\definecolor{brown}{rgb}{0.5, 0.5, 0}
\definecolor{dkgreen}{rgb}{0,0.5,0}
\definecolor{orange}{rgb}{1,.5,0}
\definecolor{dandelion}{cmyk}{0,0.29,0.84,0}

\setacronymstyle{long-short}
\newacronym{AnEn}{AnEn}{Analog Ensemble}
\newacronym{NWP}{NWP}{Numerical Weather Prediction}
\newacronym{NAM}{NAM}{North American Mesoscale Forecast System}
\newacronym{PDF}{PDF}{Probability Distribution Function}
\newacronym{CRPS}{CRPS}{Continuous Ranked Probability Score}
\newacronym{NCAR}{NCAR}{National Center of Atmospheric Research}
\newacronym{PV}{PV}{Photovoltaic}
\newacronym{WRF}{WRF}{Weather Research and Forecast model}
\newacronym{METAR}{METAR}{Meteorological Terminal Aviation Routine}
\newacronym{AUA}{AUA}{Adaptive Unstructured Analog}

\newif\ifdraft

\ifdraft
 \newcommand{\jhanote}[1]{ {\textcolor{red} { ***SJ\@: #1 }}}
 \newcommand{\mtnote}[1]{ {\textcolor{orange} { ***MT\@: #1 }}}
 \newcommand{\mlnote}[1]{ {\textcolor{blue} { ***ML\@: #1 }}}
 \newcommand{\vbnote}[1]{ {\textcolor{purple} { ***VB\@: #1 }}}
\else
 \newcommand{\jhanote}[1]{}
 \newcommand{\mtnote}[1]{}
 \newcommand{\mlnote}[1]{}
 \newcommand{\vbnote}[1]{}
\fi

\newcommand{\up}{\vspace*{-1em}}

\newcommand{\B}[1]{\textbf{#1}\xspace}

\lstdefinestyle{myListing}{
  frame=single,
  backgroundcolor=\color{listinggray},
  language=C,
  basicstyle=\ttfamily \footnotesize,
  breakautoindent=true,
  breaklines=true
  tabsize=2,
  captionpos=b,
  aboveskip=0em,
  belowskip=-2em,
}

\lstdefinestyle{myPythonListing}{
  frame=single,
  backgroundcolor=\color{listinggray},
  language=Python,
  basicstyle=\ttfamily \scriptsize,
  breakautoindent=true,
  breaklines=true
  tabsize=2,
  captionpos=b,
}

\ifpdf
\DeclareGraphicsExtensions{.pdf, .jpg, .tif}
\else
\DeclareGraphicsExtensions{.ps,  .eps, .jpg}
\fi

\begin{document}

\title{Harnessing the Power of Many: Extensible Toolkit for Scalable Ensemble Applications}

\author{
    \IEEEauthorblockN{
        Vivek Balasubramanian\IEEEauthorrefmark{1}\IEEEauthorrefmark{5},
        Matteo Turilli \IEEEauthorrefmark{1}\IEEEauthorrefmark{5},
        Weiming Hu\IEEEauthorrefmark{2},
        Matthieu Lefebvre\IEEEauthorrefmark{3}, 
        Wenjie Lei\IEEEauthorrefmark{3}, \\
        Ryan Modrak\IEEEauthorrefmark{3},
        Guido Cervone\IEEEauthorrefmark{2}, 
        Jeroen Tromp\IEEEauthorrefmark{3} and
        Shantenu Jha\IEEEauthorrefmark{1}\IEEEauthorrefmark{4},
    }
    \IEEEauthorblockA{
        \IEEEauthorrefmark{1}ECE, Rutgers University
    }
    \IEEEauthorblockA{
        \IEEEauthorrefmark{2} Penn State University
    }
    \IEEEauthorblockA{
        \IEEEauthorrefmark{3} Princeton University
    }
    \IEEEauthorblockA{
        \IEEEauthorrefmark{4} Brookhaven National Laboratory
    }
    \IEEEauthorblockA{
        \IEEEauthorrefmark{5} Contributed Equally
    }
}

\maketitle
\begin{abstract}
Many scientific problems require multiple distinct computational tasks to be
executed in order to achieve a desired solution. We introduce the Ensemble
Toolkit (EnTK) to address the challenges of scale, diversity and reliability
they pose. We describe the design and implementation of EnTK, characterize
its performance and integrate it with two exemplar use cases: seismic
inversion and adaptive analog ensembles. We perform nine experiments,
characterizing EnTK overheads, strong and weak scalability, and the
performance of the two use case implementations, at scale and on production
infrastructures. We show how EnTK meets the following general requirements:
(i) implementing dedicated abstractions to support the description and
execution of ensemble applications; (ii) support for execution on
heterogeneous computing infrastructures; (iii) efficient scalability up to
\(O(10^4)\) tasks; and (iv) task-level fault tolerance. We discuss novel
computational capabilities that EnTK enables and the scientific advantages
arising thereof. We propose EnTK as an important addition to the suite of
tools in support of production scientific computing.
\end{abstract}

\section{Introduction}\label{sec:intro}
Traditionally, advances in high-performance scientific computing have focused
on the scale, performance and optimization of an application with a large but
single task, and less on applications comprised of multiple tasks. However,
many scientific problems are expressed as applications that require multiple
distinct computational tasks to be executed in order to achieve a desired
solution.

``Task'' is used to represent processes at different scales and granularity.
In this paper, a computational task is a generalized term for a stand-alone
process that has well defined input, output, termination criteria, and
dedicated resources. Specifically, a task is used to represent an independent
simulation or data processing analysis, running on one or more nodes of a
high-performance computing (HPC) machine.

When the collective outcome of a set of tasks is of importance, this set is
defined to be an ensemble. Individual tasks within the set might be coupled
or uncoupled. When coupled, tasks might have global (synchronous) or local
(asynchronous) exchanges, and regular or irregular communication. This is in
contrast to traditional parameter sweeps, or high-throughput computing (HTC)
applications, where tasks are typically identical, uncoupled, idempotent and
can be executed in any order. Individual tasks within the ensemble may also
vary in their type, executable, and resource requirements. 


The number and type of applications that can be formulated as ensembles is
vast and span many scientific domains. Some scientific problems that have
traditionally been expressed as a single computational task must be
reformulated using ensembles so as to overcome limitations of single task
execution~\cite{cheatham2015impact}. For example, in biomolecular sciences,
due to the end of Dennard scaling, and thus limited strong scaling of
individual MD tasks, there has been a shift from running single long running
tasks towards multiple shorter running tasks, as evidenced by a proliferation
of ensemble-based algorithms~\cite{chodera2014markov,noe2009constructing}.

The execution of an ensemble on HPC machines presents three main challenges:
(1) encoding scientific problems into algorithms that are amenable to
distributed and coordinated solution; (2) sizing, acquiring, and managing
resources for the execution; and (3) managing the execution of the ensemble.
Encoding scientific problems into ensembles requires describing tasks with
heterogeneous properties, specifying whether and how tasks are grouped into
partitions of the ensemble, and defining an ordering among tasks and
partitions. Sizing resources for the ensemble depends on calculating the
resources needed by each task and those needed by the set of tasks that can
be executed concurrently.

Often, there is a friction between the resource requirements of an ensemble
and the traditional resource management of HPC machines. Each task of the
ensemble has to be queued onto a HPC machine, incurring a long queue waiting
time that adds up to the total time to completion of the ensemble. Usually,
at least one compute node of the HPC machine has to be requested for each
task, and often for at least one hour, even when tasks may require fewer
resources for a shorter duration. Finally, distributing the execution of an
ensemble requires tailored coordination and communication infrastructure and
protocols, not made readily available to the user via the HPC software
provision. These factors make using HPC resources for ensemble applications 
challenging, when not unfeasible.

In response to these challenges and requirements, the growing importance of
ensemble-based applications in scientific HPC, and the absence of middleware
providing scalable, extensible and general solutions, we have designed and
implemented the Ensemble Toolkit (EnTK). EnTK promotes ensembles to a
high-level programming abstraction, providing specific interfaces and
execution models for ensemble-based applications. EnTK is engineered for
scale and a diversity of computing platforms and runtime systems, and it is
agnostic of the size, type and coupling of the tasks comprising the ensemble.

EnTK adheres to the building blocks approach for the design, development and
integration of middleware~\cite{bb1,bb2}. This approach advocates a
sustainable ecosystem of software components from which tailored workflow
systems can be composed, as opposed to having to fit workflows to
pre-existing frameworks. The building blocks approach overcomes the limited
flexibility of monolithic workflow systems by enabling composability and
extensibility, and thereby supporting the wide range of workflow
requirements. As circumstantial evidence, EnTK has been used to develop
several diverse domain-specific workflow systems~\cite{bb1}.

This paper offer four main contributions: (1) a description of the design,
architecture and implementation of EnTK\@ (\S\ref{sec:entk}); (2) a
characterization of EnTK overheads on different HPC computing infrastructures
(CI) for a variety of task types (\S\ref{ssec:entk_validation}); (3) an
analysis of EnTK weak and strong scaling on a leadership-class HPC CI
(\S\ref{ssec:entk-scaling}); and (4) the support of two ensemble-based
scientific applications with different characteristics and requirements
(\S\ref{ssec:exp_use_cases}). This shows that EnTK can support diverse types
of ensemble applications, at scale and on several HPC CIs, introducing
acceptable overheads. As such, we propose EnTK as an important addition to
the suite of tools in support of production scientific computing.

\section{Ensemble Toolkit (ENTK)}\label{sec:entk}
The design and implementation of EnTK are iterative and driven by use cases.
Use cases span several scientific domains, including Biomolecular Sciences,
Material Sciences, and Earth Sciences. Users and developers collaborate to
elicit requirements and rapid prototyping. EnTK is loosely specified in UML,
validated against its requirements and characterized via a profiler. Jenkins
and Travis are used for continuous integration and automated testing.
Documentation and code are managed and made available via a GitHub
repository~\cite{entk-repo}.

\subsection{Requirements}\label{ssec:requirements}

As seen in~\S\ref{sec:intro}, the space of ensemble applications (hereafter
simply `applications') is vast, and thus there is a need for simple and
uniform abstractions while avoiding single-point solutions. We elicited
requirements about computing infrastructures (CIs), scale, fault-tolerance,
and usability. EnTK is required to: (1) support heterogeneous CIs; (2)
abstract the complexity of execution and resource management; and (3) be
performance independent of the type of CI\@.


The use cases motivating EnTK require execution of up to \(O(10^4)\) ensemble
members (tasks). This poses many challenges, that need to be addressed by
EnTK and a runtime system (RTS). At this scale, EnTK has to reliably enable
sustained task submission rate, tracking of executing tasks and clean
termination of tasks. The RTS has instead to integrate with suitable MPI
layers, setting up the execution environment for heterogeneous tasks,
managing their data requirements and scheduling tasks across multiple
resource partitions. Together, EnTK and RTS have to ensure full resource
utilization across the ensemble execution time.


EnTK has to be fault-tolerant at scale, i.e., when both the probability and
cost of failure increase. Currently, EnTK is required to support resubmission
of failed tasks, without application checkpointing, and restarting of failed
RTS and components. In this way, applications can be executed on multiple
attempts, without restarting completed tasks.


Usability plays an important role in the development of EnTK, as it must
support diverse programming and development skills. Special attention is
given to lowering the time to encode use cases into executable applications.

\subsection{Design}\label{ssec:design}


\subsubsection{Application Model} 

We model applications by combining 
the following user-facing constructs:

\begin{itemize}
  \item \textbf{Task:} an abstraction of a computational task that contains
  information regarding an executable, its software environment and its data
  dependences.
  \item \textbf{Stage:} a set of tasks without mutual dependences and that
  can be executed concurrently.
  \item \textbf{Pipeline:} a list of stages where any stage \(i\) can be
  executed only after stage \(i-1\) has been executed.
\end{itemize}

\begin{figure}
  \includegraphics[width=\columnwidth]{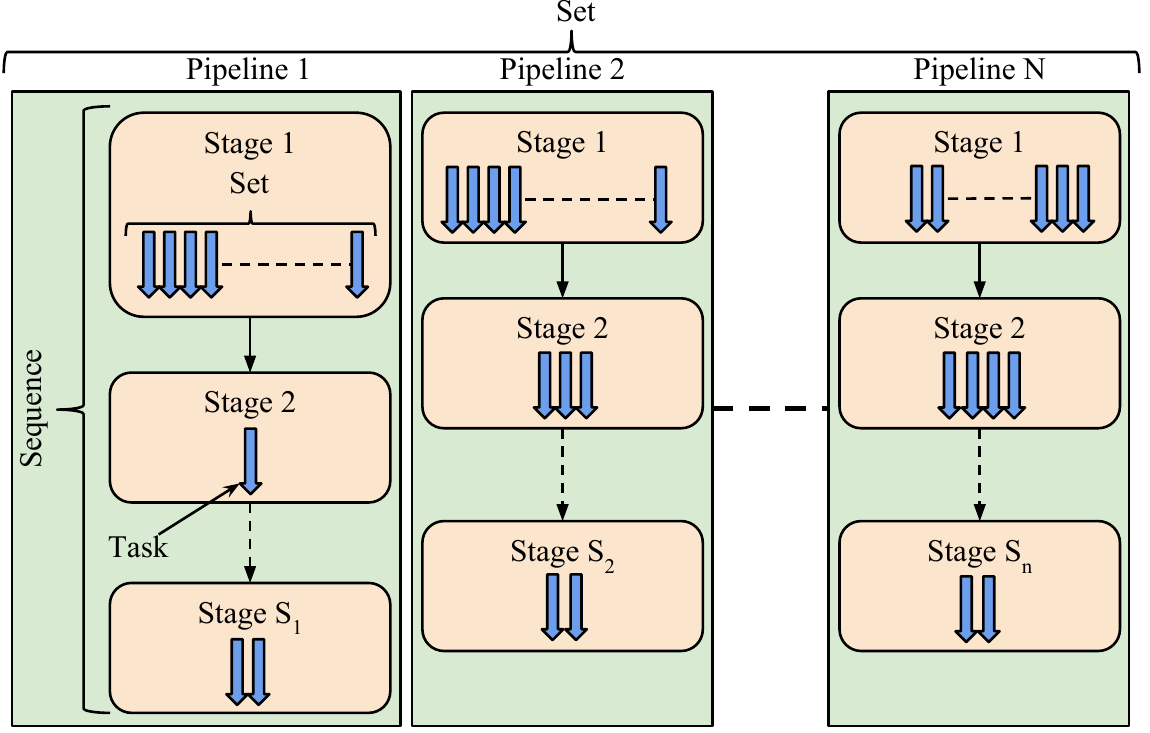}
  \caption{Diagrammatic representation of an application consisting of a set
          of pipelines with varying number of stages and
          tasks.}\label{fig:pst-model}
\up{}
\end{figure}

Figure~\ref{fig:pst-model} shows an application described with pipelines,
stages, and tasks (PST). The application consists of a set of pipelines,
where each pipeline is a list of stages, and each stage is a set of tasks.
All the pipelines can execute concurrently, all the stages of each pipeline
can execute sequentially, and all the tasks of each stage can execute
concurrently.

Note that PST descriptions can be extended to account for dependencies among
groups of pipelines in terms of lists of sets of pipelines. Further, the
specification of branches in the execution flow of applications does not
require to alter the PST semantics: Branching events can be specified as
tasks where a decision is made about the runtime flow. For example, a task
could be used to decide to skip some elements of a stage, based on some
partial results of the ongoing computation.

\subsubsection{Architecture} 

\begin{figure}
 \includegraphics[width=\columnwidth]{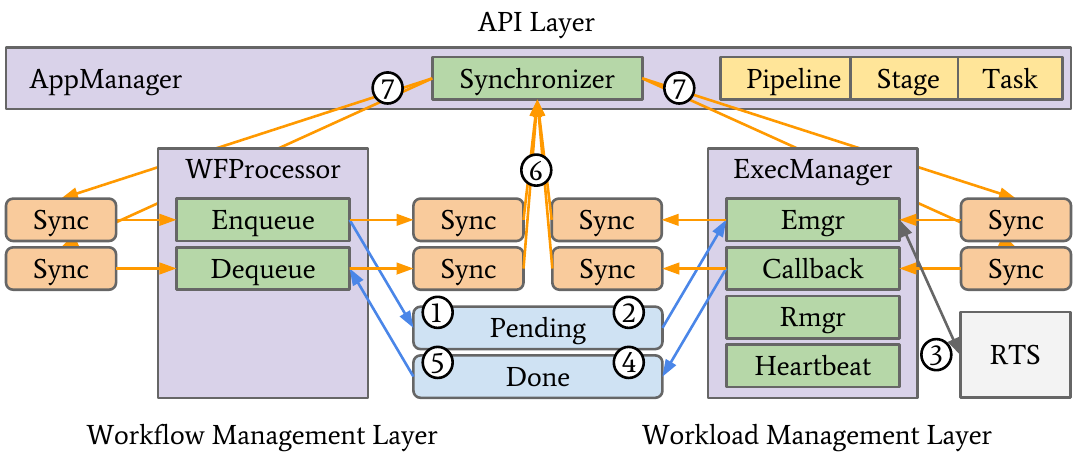}
 \caption{EnTK architecture and execution model. Components' (purple)
 subcomponents (green) use queues (blue and orange) to communicate and
 coordinate the execution of an application via a chosen RTS
 (gray).}\label{fig:entk-arch}
\up{}
\end{figure}

EnTK sits between the user and the CI, abstracting resource management and
execution management from the user. Fig.~\ref{fig:entk-arch} shows the
components (purple) and subcomponents (green) of EnTK, organized in three
layers: API, Workflow Management, and Workload Management.

The API layer enables users to codify PST descriptions. The Workflow
Management layer retrieves information from the user about available CIs,
initializes EnTK, and holds the global state of the application during
execution. The Workload Management layer acquires resources via the RTS.

The Workflow Management layer has two components: AppManager and WFProcessor.
AppManager uses the Synchronizer subcomponent to update the state of the
application at runtime. WFProcessor uses the Enqueue and Dequeue
subcomponents to queue and dequeue tasks from the Workload Management layer. The
Workload Management layer uses ExecManager and its Rmgr, Emgr, RTS Callback,
and Heartbeat subcomponents to acquire resources from CIs and execute the
application.

Another benefit of this architecture is the isolation of the RTS into a
stand-alone subsystem. This enables composability of EnTK with diverse RTS
and, depending on capabilities, multiple types of CIs. Further, EnTK assumes
the RTS to be a black box enabling fault-tolerance. When the RTS fails or
becomes unresponsive, EnTK can tear it down and bring it back, loosing only
those tasks that were in execution at the time of the RTS failure.

\subsubsection{Execution Model}

EnTK components and subcomponents communicate and coordinate for the
execution of tasks. Users describe an application via the API, instantiate
the AppManager component with information about the available CIs and then
pass the application description to AppManager for execution. AppManager
holds these descriptions and, upon initialization, creates all the queues,
spawns the Synchronizer, and instantiates the WFProcessor and ExecManager.
WFProcessor and ExecManager instantiate their own subcomponents.

Once EnTK is fully initialized, WFProcessor initiates the execution by
creating a local copy of the application description from AppManager and
tagging tasks for execution. Enqueue pushes these tasks to the Pending queue
(Fig.~\ref{fig:entk-arch}, 1). Emgr pulls tasks from the Pending queue
(Fig.~\ref{fig:entk-arch}, 2) and executes them using a RTS
(Fig.~\ref{fig:entk-arch}, 3). RTS Callback pushes tasks that have completed
execution to the Done queue (Fig.~\ref{fig:entk-arch}, 4). Dequeue pulls
completed tasks (Fig.~\ref{fig:entk-arch}, 5) and tags them as done, failed
or canceled, depending on the return code from the RTS\@.

Throughout the execution of the application, tasks, stages and pipelines
undergo multiple state transitions in both WFProcessor and ExecManager. Each
component and subcomponent synchronizes these transitions with AppManager by
pushing messages through dedicated queues (Fig.~\ref{fig:entk-arch}, 6).
AppManager pulls these messages and updates the application states.
AppManager then acknowledges the updates via dedicated queues
(Fig.~\ref{fig:entk-arch}, 7). This messaging mechanism ensures that
AppManager is always up-to-date with any state change, making it the only
stateful component of EnTK\@.

\subsubsection{Failure Model}

We consider four main sources of failure: EnTK components, RTS, CI, and task
executables. All state updates in EnTK are transactional, hence any EnTK
component that fails can be restarted at runtime without losing information
about ongoing execution. In case of full failure, EnTK can reacquire upon
restarting information about the state of the execution up to the latest
successful transaction before the failure. Information is synced on disk and
hooks are in place to use an external database.

Both the RTS and the CI are considered black boxes. Partial failures of their
subcomponents at runtime are assumed to be handled locally, not globally by
EnTK\@. Upon full failure of the RTS, EnTK assumes all the pilot resources
and the tasks undergoing execution are lost. EnTK purges any process left
over by the failed RTS, starts a new instance of the RTS, acquires new pilot
resources, and restarts executing the ensemble until completion. Users can
configure the number of times a RTS is restarted during the execution of a
single ensemble.

CI-level failures are reported to EnTK indirectly, either as failed pilots
or failed tasks. Both pilots and tasks can be restarted, up to a certain
number of times configured by the user. Failures are logged and reported to
the user at runtime for live or postmortem analysis. EnTK design enables
collection of information from both the RTS and the resources via APIs. When
RTS and CI expose information about, for example, OS-level faults,
application checkpoint, or hardware faults, EnTK can implement advanced
fault-tolerant capabilities. Currently, these capabilities are not required
by our use cases and application checkpoint is not performed by their
executables.

\subsection{Implementation}\label{ssec:implementation}

EnTK is implemented in Python, uses RabbitMQ message queuing system and the
RADICAL-Pilot (RP) runtime system. All EnTK components are implemented as
processes, and all subcomponents as threads. AppManager is the master process
spawning all the other processes. Tasks, stages and pipelines are implemented
as objects, copied among processes and threads via queues and transactions.
Synchronization among processes is achieved by message-passing via queues.

EnTK relies on RabbitMQ to manage the creation of the communication
infrastructure to transport the objects and messages among components.
RabbitMQ provides methods to increase the durability of messages in transit
and of the queues, and to acknowledge messages. Most importantly, it supports
the requirement of managing at least \(O(10^4)\) tasks concurrently.

RabbitMQ is a server-based system and requires to be installed before the
execution of EnTK\@. This adds overheads but it also offers the following
benefits: (1) producers and consumers do not need to be topology aware
because they interact only with the server; (2) messages are stored in the
server and can be recovered upon failure of EnTK components; and (3) messages
can be pushed and pulled asynchronously because data can be buffered by the
server upon production.

\subsection{Runtime System}\label{ssec:rts}



Currently, EnTK uses RADICAL-Pilot
(RP)~\cite{rp-paper-isc,review_radicalpilot} as the RTS. RP is a
runtime system designed to execute ensemble applications via pilots. Pilots
provide a multi-stage execution mechanism: Resources are acquired via a
placeholder job and subsequently used to execute the application's tasks.
When a pilot is submitted to a CI as a job, it waits in the CI's queue until
the requested resources become available. At that point, the CI's scheduler
bootstraps the job on the CI's compute nodes. RP does not attempt to `game'
the CI's scheduler: Once queued, the pilot is managed according to the CI's
policies.

\begin{figure}
 \includegraphics[width=\columnwidth]{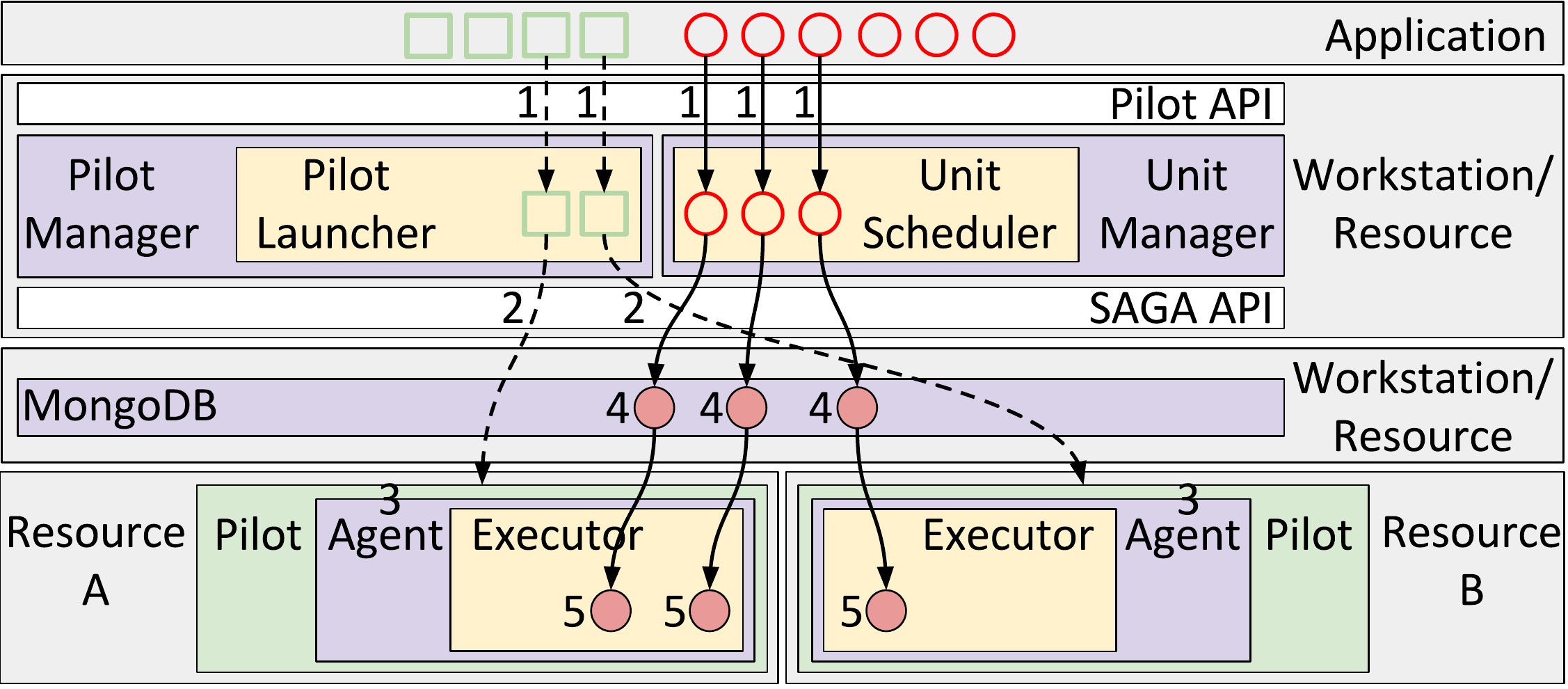}
  \caption{RADICAL-Pilot (RP) architecture and execution model. Gray:
  machines; green: pilot; purple: modules; yellow: components; red:
  tasks.}\label{fig:arch-overview}
\end{figure}

RP is a distributed system with four modules: PilotManager, UnitManager,
Agent and DB (Fig.~\ref{fig:arch-overview}, purple boxes). PilotManager,
UnitManager and Agent have multiple components (Fig.~\ref{fig:arch-overview},
yellow boxes), isolated into separate processes. Components are stateless and
some of them can be instantiated concurrently to enable RP to manage multiple
pilots and tasks at the same time. Concurrent components are coordinated via
a dedicated communication mesh, scaling throughput and enabling tolerance to
failing components.

Workloads and pilots are described via the Pilot API and passed to the RP
runtime system (Fig.~\ref{fig:arch-overview}, 1). The PilotManager submits
pilots as jobs (or virtual machines or containers) to one or more CIs via the
SAGA API (Fig.~\ref{fig:arch-overview}, 2). The SAGA API implements an
adapter for each supported type of CI, exposing uniform methods for job and
data management. Once a pilot becomes active on a CI, it bootstraps the Agent
module (Fig.~\ref{fig:arch-overview}, 3). The UnitManager schedules each task
to an Agent (Fig.~\ref{fig:arch-overview}, 4) via a queue on a MongoDB
instance. Each Agent pulls its tasks from the DB module
(Fig.~\ref{fig:arch-overview}, 5), scheduling them on the Executor. The
Executor sets up the task's execution environment and then spawns the task
for execution.


When required, the input data of a task are either pushed to the Agent or
pulled from the Agent, depending on data locality and sharing requirements.
Similarly, the output data of the task are staged out by the Agent and
UnitManager to a specified destination, e.g., a filesystem accessible by the
Agent or the user workstation. Both input and output staging are optional,
depending on the requirements of the tasks. The actual file transfers are
enacted via SAGA, and currently support \texttt{(gsi)-scp},
\texttt{(gsi)-sftp}, \texttt{Globus Online}, and local and shared filesystem
operations via \texttt{cp}. Consequently, the size of the data along with
network bandwidth and latency or filesystem performance determine the data
staging durations and are independent of the performance of the RTS\@.



\section{Use cases}\label{sec:use_cases}
To help understand the initial scope and design of EnTK, we describe two
motivating use cases, focusing on their computational and functional
requirements.


\subsection{Seismic Inversion}\label{ssec:seismic_inversion}

Inversion of full-waveform, wide-bandwidth seismic data~\cite{Virieux2009} is
one of the most powerful tomographic technique to study the Earth's interior.
Scaling this technique is challenging, mostly because of the amount of
computational resources and human labor it needs. These challenges require a
more automated approach to the management and execution of the workflow, such
as the one implemented by the EnTK\@.

\begin{figure}
   \centering
   \includegraphics[width=\columnwidth]{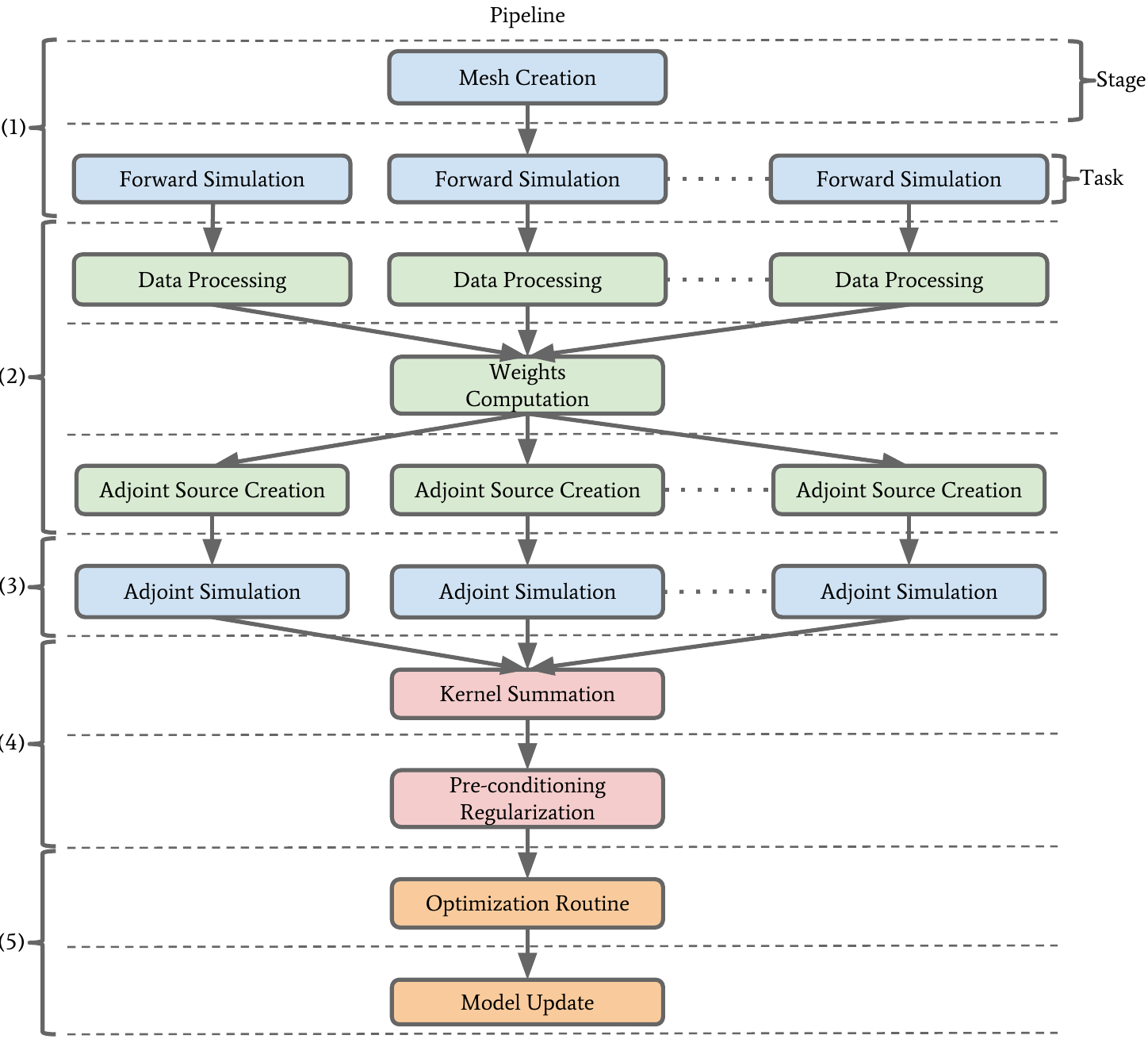} 
   \caption{Simplified seismic tomography workflow encoded into the PST 
   model.}\label{fig:use_case_tomo_wf}
\up{}
\end{figure}

Figure~\ref{fig:use_case_tomo_wf} shows a high level view of the workflow we
use to perform seismic tomography. We record seismic data (i.e., seismograms)
as time series of a physical quantity, like displacement, velocity,
acceleration or pressure. Our goal is to iteratively minimize differences
between observed and corresponding synthetic data through a pre-defined
misfit function. As the adjoint-based optimization procedure is carried on
and the data misfit decreases, the model gets closer to reality.

We run the workflow of Figure~\ref{fig:use_case_tomo_wf} in production,
assimilating data from about 1,000 earthquakes. Forward (1) and adjoint (3)
simulations are the most computationally expensive parts of the workflow,
each running on 384 GPUs for a total of 10~millions core-hours per iteration.
Data processing (2) is relatively computationally inexpensive, utilizing
about 48,000 core-hours in each iteration. Post-processing (4) takes about
10,000 core-hours while optimization (5) takes about 1 million core-hours.

Currently, each part of the workflow relies on a Python-based proto-workflow
management system. However, scaling to higher resolutions and assimilating
data from 6,000 earthquakes requires more automation to ensure reliability,
minimize errors at the user level and lower the overall time to solution.
Further, we need to interleave simulation tasks with data-processing tasks,
each requiring respectively leadership-scale systems and moderately sized
clusters. During the workflow execution, we need to save between 0.15 to
1.5GB per seismogram.

Ensemble-based applications are particularly well-suited to encode the
seismic tomography workflow. In EnTK, we can describe the simulation and
analysis phases as stages of a pipeline, avoiding to use dedicated MPI
application to execute multiple simulation concurrently. EnTK and RP also
allow the execution of the ensemble of simulations with a varying degree of
concurrency and sequentiality, without requiring specific coding. EnTK offers
automation and fault-tolerance avoiding the overheads we experienced with
full-fledged workflow systems. We encode data dependencies and staging
directives via the EnTK API, enabling data management at runtime on per-task,
stage, and pipeline basis.

\subsection{High Resolution Meteorological Probabilistic
Forecasts}\label{ssec:anen}


We implemented the \gls{AnEn}~\cite{anen} methodology to generate
high-resolution, probabilistic forecasts for environmental variables like
temperature or cloud cover. We used relationships between current and past
forecasts from the \gls{WRF} data to generate an analog ensemble for a given
time and location. Our implementation finds the most similar historical
forecasts, based on a similarity metric. The observations associated with the
most similar past forecasts are used as analogs.

We implemented a dynamic iterative search process, named the \gls{AUA}
algorithm, which generates analogs at specific geographical locations, and
interpolates the analogs using an unstructured grid. In this way, we avoid
computing analogs at every available location, noting that for some output
variables, such as temperature, the highest resolution of the analogs is
required only at specific regions, where drastic gradient changes occur.


The \gls{AUA}  algorithm is iterative, and at each iteration it performs a
variable number of operations. EnTK addresses the resource management
challenges arising from such variations by allocating diverse amount of
computing resources, depending on the size of the search space required to
achieve the desired prediction accuracy.

\begin{figure}[t]
    \centering
    \includegraphics[width=\columnwidth]{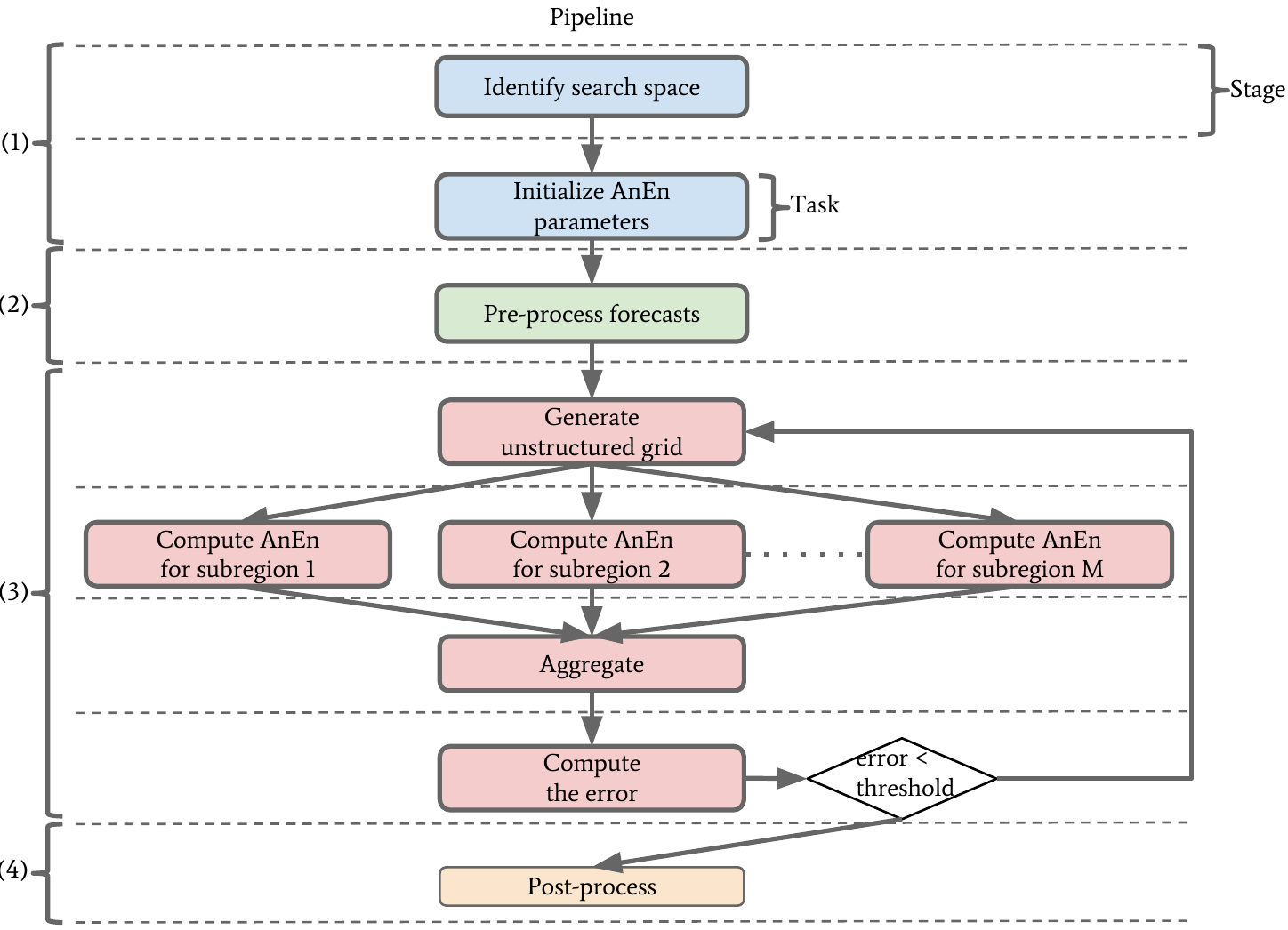}
    \caption{Adaptive unstructured analog algorithm
    workflow encoded into the PST model.}\label{fig:psu-use-case}
\up{}
\end{figure}

Figure~\ref{fig:psu-use-case} shows the workflow of the \gls{AUA} algorithm.
The initialization step specifies the search space and the test space, and
sets up starting parameters for the \gls{AnEn}. The preprocessing step
generates preparatory data for the subsequent steps. The largest amount of
computation occurs in the iterative computation step where analogs are
computed and aggregated multiple times until the available resources are
exhausted, or the prediction error is below a given threshold. The post-
processing task interpolates the analogs to generate the forecast solution.

We drove the development and assessed the suitability of EnTK for analog
computation by testing the \gls{AnEn} method with dataset including forecast
predictions for 13 variables (e.g., wind speed, precipitation, pressure,
etc.) for the years 2015 and 2016. Data for both analysis and forecasts are
from the \gls{NAM} maintained at the \gls{NCAR}.

\section{Experiments}\label{sec:experiments}
We perform experiments to characterize the overheads of EnTK and its weak and
strong scaling performance. We then measure the overheads of EnTK when
executing, at scale, the implementation of the two use cases described
in~\S\ref{sec:use_cases}.


We use four applications in our experiments: Sleep,
Gromacs~\cite{abraham2015gromacs}, Specfem~\cite{komatitsch2002spectral}, and
Canalogs~\cite{alessandrini2015analog}. Sleep and Gromacs enable control of
the duration of task execution and to compare EnTK overheads across task
executables. Gromacs and the NTL9 protein serve as workload for EnTK weak and
strong scalability, while Specfem and Canalogs are required by the two use
cases of this paper.


We perform our experiments on four CIs: XSEDE SuperMIC, Stampede, Comet and
ORNL Titan. We use four resources for the characterization of EnTK overheads;
Titan for the characterization of the scalability; and Titan and SuperMIC for
the use case applications. 


\subsection{Characterization of EnTK Performance}\label{ssec:entk_validation}

We use a prototype of EnTK to benchmark its performance, providing a
reference hardware configuration to support execution of up to \(O(10^6)\)
tasks. We then perform four experiments to characterize the overheads of
EnTK\@.

\subsubsection{Performance of EnTK Prototype}

We prototyped the most computationally expensive functionality of EnTK to
instantiate multiple producers and consumers of tasks. Each producer pushes
tasks into RabbitMQ queues and each consumer pulls tasks from these queues,
passing them to an empty RTS module.
We benchmarked configurations with \(10^6\) tasks and a different number of
producers, consumers, and queues, measuring: producers and consumers time;
total execution time; base memory consumption when the components are
instantiated; and peak memory consumption during the execution.

\begin{figure}[t]
  \includegraphics[width=\columnwidth]{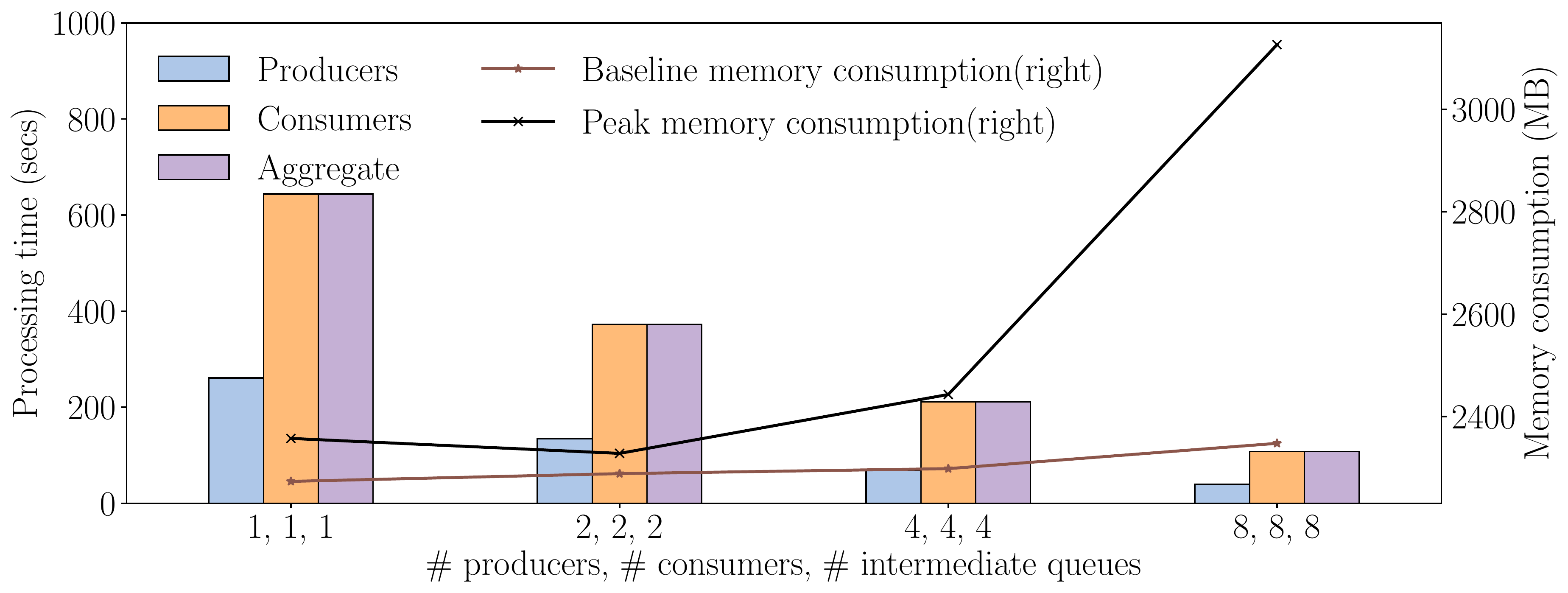}
  \caption{Execution time and memory consumed by EnTK prototype with multiple
  producers and consumers and \(10^6\) tasks.}\label{fig:entk_prototype_perf}
\up{}
\end{figure}

Fig.~\ref{fig:entk_prototype_perf} shows that tuning of the prototype can
reduce the processing time linearly, at the cost of increased memory usage.
Eight producers and consumers require 107 seconds to process \(10^6\) tasks,
with a peak memory consumption of 3,126MB\@. Uneven distributions of
producers and consumers resulted in lower efficiencies than when using even
distributions.

The execution model of EnTK can be tuned on the basis of this benchmark,
workload requirements, and hardware capabilities. This benchmark shows that
the performance of the core functionality of EnTK depends on the number of
tasks that are processed concurrently. This has relevant implications for the
understanding of EnTK overheads and scalability.

\subsubsection{Overheads, Data Staging and Task Execution Time}


We characterize EnTK overhead against four parameters that are likely to vary
among applications: Task executable; task duration; CI on which the
application is executed; and structure of the application, i.e., the way in
which tasks are grouped into stages and stages into pipelines. We measured
the overheads that dominate EnTK and RTS runtime alongside the total task
execution time and, when required, the total data staging time:

\begin{itemize}
    \item \textbf{EnTK Setup Overhead}: Time taken to setup the messaging
    infrastructure, instantiate components and subcomponents, and validate
    application and resource descriptions.
    \item \textbf{EnTK Management Overhead}: Time taken to process the
    application, translate tasks from and to RTS-specific objects, and
    communicate pipelines, stages, tasks and control messages.
    \item \textbf{EnTK Tear-Down Overhead}: Time taken to cancel all EnTK
    components and subcomponents, and shutdown the messaging infrastructure.
    \item \textbf{RTS Overhead}: Time taken by the RTS to submit and manage
    the execution of the tasks.
    \item \textbf{RTS Tear-Down Overhead}: Time taken by the RTS to cancel
    its components and to shutdown.
    \item \textbf{Data Staging Time}: Time taken to copy data between
    tasks using the functionality available on the resource (in this case,
    the Unix POSIX \texttt{cp} command).
    \item \textbf{Task Execution Time}: Time taken by the task executables to
    run on the CI\@.
\end{itemize}

\begin{figure*}[!ht]
    \centering
    \includegraphics[width=\textwidth]{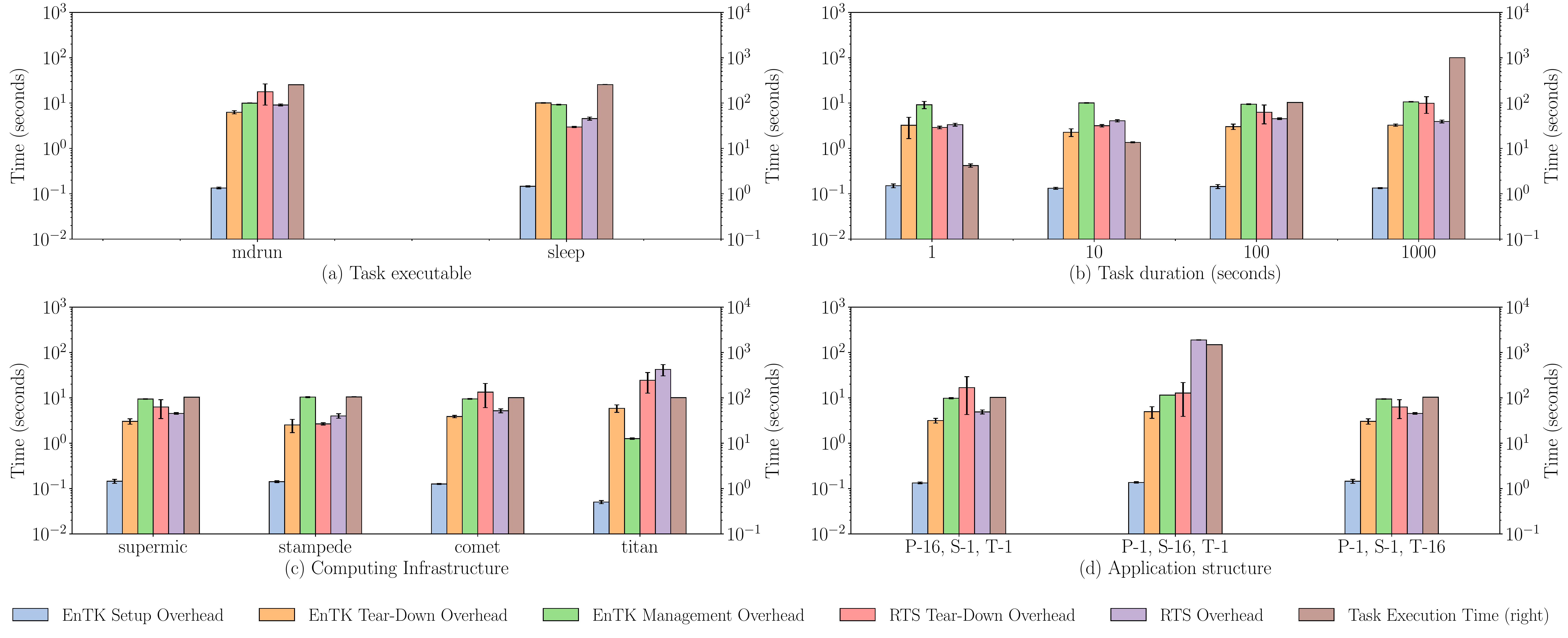}
    \caption{Overheads and Task Execution Time as function of (a) Task
             Executable (Experiment 1), (b) Task Duration (Experiment 2) (c)
             Computing Infrastructure (Experiment 3) (d) Application
             Structure (Experiment 4).}\label{fig:overheads}
\end{figure*}

We designed four experiments (Table~\ref{tab:experiments}) to characterize
the overheads added by EnTK and the RP RTS to the time taken to execute an
application, excluding the time taken by the resources to become available.
These experiments execute applications with different task executable
(Experiment 1, Fig.~\ref{fig:overheads}a); task duration (Experiment 2,
Fig.~\ref{fig:overheads}b); CI (Experiment 3, Fig.~\ref{fig:overheads}c); and
application structure, i.e., the number of pipelines, stages and tasks per
application (Experiment 4, Fig.~\ref{fig:overheads}d).

\begin{table*}
    \caption{Parameters of the experiments plotted in
    Figure~\ref{fig:overheads}.}\label{tab:experiments}
    \centering
    \resizebox{\textwidth}{!}{
    \begin{tabular}{l            
                    l            
                    l            
                    l            
                    l            
                    l            
                    }
    \toprule
    \B{ID}                            &  
    \B{Computing Infrastructure (CI)} &  
    \B{Pipeline, Stage, Task}         &  
    \B{Executable}                    &  
    \B{Task Duration}                 &  
    \B{Data}                          \\ 
    \midrule
    \B{1}                             &  
    SuperMIC                          &  
    (1,1,16)                          &  
    \texttt{mdrun}, \texttt{sleep}    &  
    300s                              &  
    TDB                               \\ 
    \B{2}                             &  
    SuperMIC                          &  
    (1,1,16)                          &  
    \texttt{sleep}                    &  
    1s, 10s, 100s, 1,000s              &  
    None                              \\ 
    \B{3}                             &  
    SuperMIC, Stampede, Comet, Titan  &  
    (1,1,16)                          &  
    \texttt{sleep}                    &  
    100s                              &  
    None                              \\ 
    \B{4}                             &  
    SuperMIC                          &  
    (16,1,1), (1,16,1), (1,1,16)      &  
    \texttt{sleep}                    &  
    100s                              &  
    None                              \\ 
    \bottomrule
    \end{tabular}
    }
\up{}
\end{table*}

Fig.~\ref{fig:overheads} shows that EnTK Setup Overhead is \(\approx\)0.1s
across Experiment 1, 2, 4, and \(\approx\)0.05s for Titan in Experiment 3. We
attribute this difference to the host from which EnTK was executed. All the
experiments on XSEDE machines were performed from the same virtual machine
(VM) hosted at TACC, while experiments on Titan had to be performed from an
ORNL login node. The ORNL login nodes have faster memory and CPU than the
VM\@.

Fig.~\ref{fig:overheads} shows a similar behavior between EnTK Setup Overhead
and EnTK Management Overhead. EnTK Management Overhead measures
\(\approx\)10s on all the runs but those performed on Titan where it measures
\(\approx\)3s. Also in this case, we attribute this difference to the
performance of the VM and login nodes from which EnTK was executed.

In Fig.~\ref{fig:overheads}, EnTK Tear-Down Overhead and RTS Tear-Down
Overhead vary across all four experiments with values between \(\approx\)1
and \(\approx\)10s for EnTK Tear-Down Overhead and \(\approx\)3 and
\(\approx\)80s for RTS Tear-Down Overhead. We attribute these variations to
the time taken by Python to terminate processes and threads. The higher
values of RTS Tear-Down Overhead are expected as RP uses significantly more
processes and threads than EnTK\@.

We explain the variations of RTS Overhead in Fig.~\ref{fig:overheads} by
noticing that, at runtime, RP initiates communications between the CI and a
remote database, and reads and writes to the shared file system of the CI to
create the execution environment of each task. Further, RP uses third party
tools to distribute the execution of tasks across compute nodes. A detailed
analysis of the interplay among network latency, I/O performance, and the
performance of third party tools and libraries is beyond the scope of this
paper. This is consistent with EnTK design: the RTS (RP in this case) is
assumed to be a black box.


Fig.~\ref{fig:overheads} shows that for tasks executing more than 1s, RP
overheads have little impact on Task Execution time: As per experiment
design, executables of Experiment 1 (Fig.~\ref{fig:overheads}a) run for
\(\approx\)300s and those of Experiment 3 (Fig.~\ref{fig:overheads}c) for
\(\approx\)100s on all four CIs. In Experiment 2 (Fig.~\ref{fig:overheads}b),
tasks set to run for 1s, run for \(\approx\)5s due to RP overhead but tasks
set to run for 10s, 100s, and 1,000s run in about that amount of time. In
Experiment 4 (Fig.~\ref{fig:overheads}c), for runs with 16 pipelines and 16
tasks, all the tasks execute concurrently and hence Task Execution Time is
\(\approx\)100s. However, with 16 stages, tasks execute sequentially,
resulting in Task Execution Time of \(\approx\)1,600s.

EnTK setup, management, and tear-down overheads vary minimally with the four
parameters of task execution we measured. Setup and management overheads
depend on the memory and CPU performance of the host on which EnTK is
executed, while the tear-down overhead on the Python version utilized. This
validates EnTK design and implementation against its requirements: EnTK can
be used in various scientific domains, with different task executables, and
across heterogeneous CIs.


In absolute terms, EnTK overheads are between \(\approx\)10 and 20 seconds
but Experiment 3 shows that these overheads can be reduced by running EnTK on
a host with better performance. RP RTS shows overheads up to \(\approx\)80s,
limiting its utilization to applications with at least minutes-long tasks.
These limitations are mostly due to the use of Python and its process and
thread termination time: EnTK and RP should be coded, at least partially, in
a different language to manage the execution of applications of 
tasks that are O(1) seconds.

\subsection{Scalability}\label{ssec:entk-scaling}

We perform two experiments to characterize weak and strong scalability of
EnTK\@. As with Experiment 1--4, we measure and compare all overheads, Data
Staging Time and Task Execution Time. Weak scaling relates these measures
to the amount of concurrency used to execute the application's tasks; Strong
scaling to the amount of serialization. 


\subsubsection{Weak scalability}

To investigate weak scaling,
we run four applications on Titan, each with 1 pipeline, 1 stage per
pipeline, and 512, 1,024, 2,048, or 4,096 tasks per stage. Each task
executable is Gromacs \texttt{mdrun}, configured to use 1 core for
\(\approx\)600 seconds. The number of acquired cores is equal to the number
of the application's tasks. Each task requires 4 input files: 3 soft links of
130B each and 1 file of 550KB\@.


\begin{figure}[t]
  \includegraphics[width=\columnwidth]{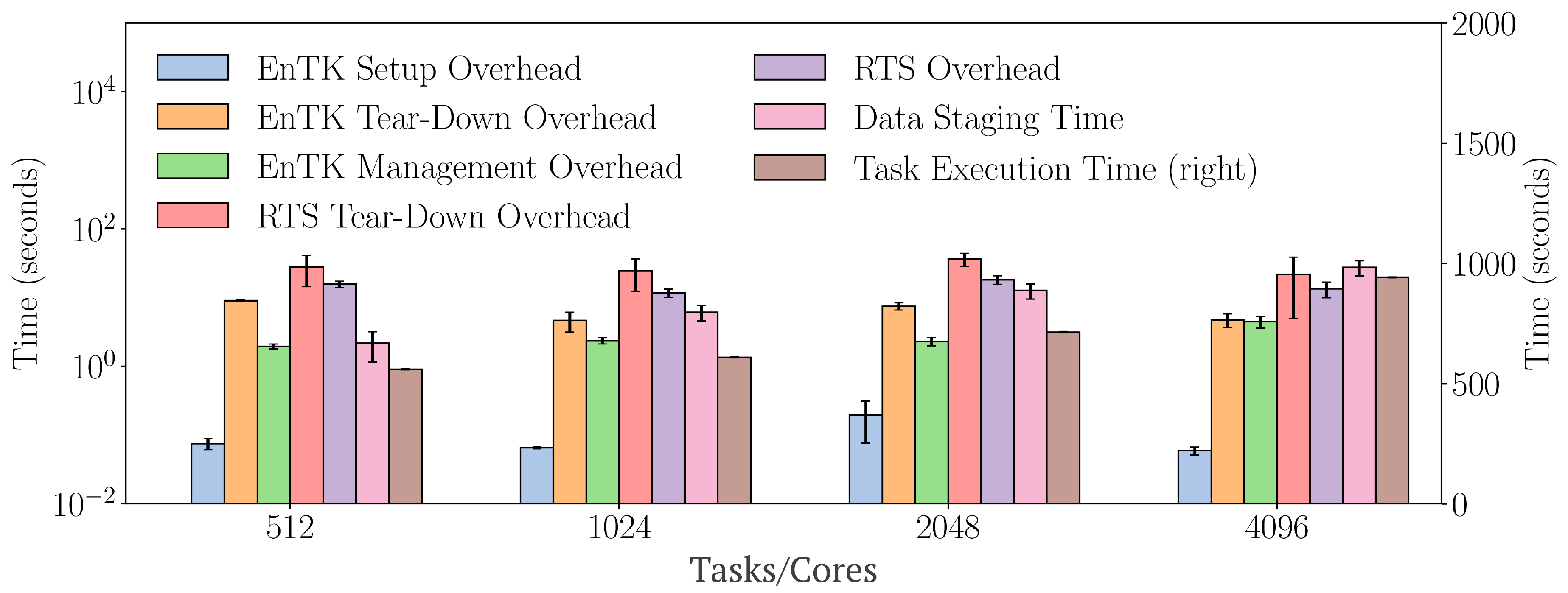}
  \caption{Weak scalability on Titan: 512, 1,024, 2,048, and 4,096
  1-core tasks executed on the same amount of cores. 
  }\label{fig:weak_scaling}
\up{}
\end{figure}

Fig.~\ref{fig:weak_scaling} (right axis) shows that Task Execution Time
increases gradually and therefore does not have ideal weak scaling. Analysis
of the RTS profiles shows that this behavior is due to delays in the Executor
module of the RTS Agent and, specifically, in the current implementation of
the Agent scheduler and the ORTE distributed virtual machine of OpenMPI\@.
Ref.~\cite {rp-paper-isc} characterizes these delays and their causes.

EnTK Management Overhead remains almost constant till 2,048 tasks as the
number of tasks are too small to cause a variation. The overhead, then,
increases between 2,048 and 4,096 tasks: With the increase of the number of
concurrent tasks, EnTK requires more resources and starts to strain the
resources of the host on which it is executed. The other EnTK and RTS
overheads appear to be consistent with those already noted in Experiments 1--4.


EnTK neither controls, nor contributes to Data Staging time. Data staging is
performed by the RP RTS that, in this experiment, creates 1 directory for
each task, writing 3 soft links and copying 1 file within it for a total of
\(\approx\)1MB\@. RP uses Unix commands to perform these operations on the
OLCF Lustre filesystem. By default, RP is configured with 1 stager and hence
files are staged sequentially. Multiple staging workers can be used to
parallelize data staging but trade offs with the filesystem performance must
be taken into account.

Data Staging time grows linearly with the number of tasks executed: from
\(\approx\)11s for 512 tasks to \(\approx\)88s for 4,096 tasks. As this time
mostly depends on the performance of Lustre, a less linear behavior is
expected with larger (amount of) files.

\subsubsection{Strong scalability}


To investigate strong scaling, 
we run four applications on Titan, each with 1 pipeline, 1 stage per
pipeline, 8,192 tasks per stage and a total of 1,024, 2,048 or 4,096 cores.
Each task executable is Gromacs \texttt{mdrun}, configured to use 1 core for
\(\approx\)600 seconds. In this way, we execute at least 2 generations, each
with 4,096 tasks, within the 2 hours walltime imposed by Titan's queuing
policies. Data staging is as in the weak scalability experiment.


\begin{figure}[t]
  \includegraphics[width=\columnwidth]{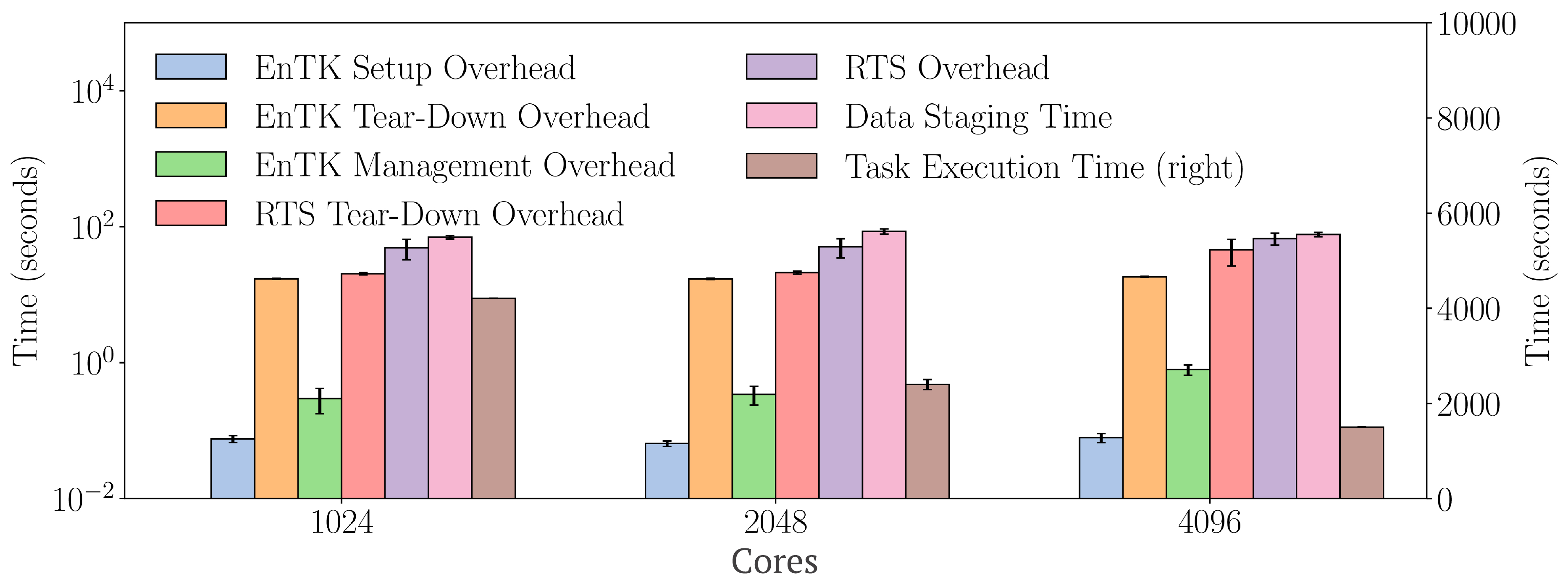}
  \caption{Strong scalability on Titan: 8,192 1-core tasks
  are executed on 1,024, 2,048 and 4,096 cores. 
  }\label{fig:strong_scaling}
\end{figure}

Fig.~\ref{fig:strong_scaling} shows that Task Execution Time reduces linearly
with increase in the number of cores. The availability of more resources for
the fixed number of tasks explains this linear reduction in the Task
Execution Time. EnTK Management Overhead is \(\approx\)1s,
confirming what already observed in the previous experiments. 

All the other overheads and Data Staging Time remain constant across the
experiment runs. This suggests that both EnTK and RP overheads mostly depend
on the number of managed tasks, not on the size of the pilot on which they
are executed. This is confirmed for RP in Ref.~\cite{rp-paper-isc}.

Fig.~\ref{fig:entk_prototype_perf} shows that EnTK can be configured to
execute \(10^{6}\) tasks in less than 200 seconds and consuming less than 4GB
of memory. Extrapolating and accounting for a faster CPU on Titan's login
nodes, EnTK should manage enough tasks to fill all of Titan cores with an
overhead of less than 20 seconds.

\subsection{Use Cases at Scale}\label{ssec:exp_use_cases}

We implement and execute at scale the most computationally intensive and
fault-prone step of the tomography workflow, and the full adaptive analog
workflow (\S\ref{sec:use_cases}) with EnTK.

\subsubsection{Seismic inversion}\label{ssec:exp_seismic}

We use EnTK to encode the forward simulations of the seismic tomography
workflow described in~\S\ref{ssec:seismic_inversion} and depicted in
Fig.~\ref{fig:use_case_tomo_wf}. These simulations account for more than 90\%
of the computation time of the workflow, requiring 384 nodes for each
earthquake simulation, and 40MB of input data each. When earthquakes are
concurrently simulated, they require a sizable portion of Titan and incur a
high rate of failures. Without EnTK, these failures result in manual
resubmission of computations, adding a significant overhead due to queue wait
time on user intervention.

We characterize the scalability of forward simulations with EnTK by running
experiments with a varying number of tasks, where each task uses 384
nodes/6,144 cores to forward simulate one earthquake. Understanding this
scaling behavior contributes to optimize the execution of the whole workflow,
both by limiting failure and enabling fault-tolerance without manual
intervention. Ultimately, this will result in an increase of the overall
efficiency of resource utilization and in a reduction of the time to
completion.

The current implementation of forward simulations causes heavy I/O on a
shared file system (\S\ref{ssec:seismic_inversion}). This overloads the file
system, inducing crashes or requiring termination of the simulations. EnTK
and RP utilize pilots to sequentialize a subset of the simulations, reducing
the concurrency of their execution and without having to go through Titan's
queue multiple times. This is done by reducing the number of cores and
increasing the walltime requested for the pilot.

\begin{figure}[t]
  \includegraphics[width=\columnwidth]{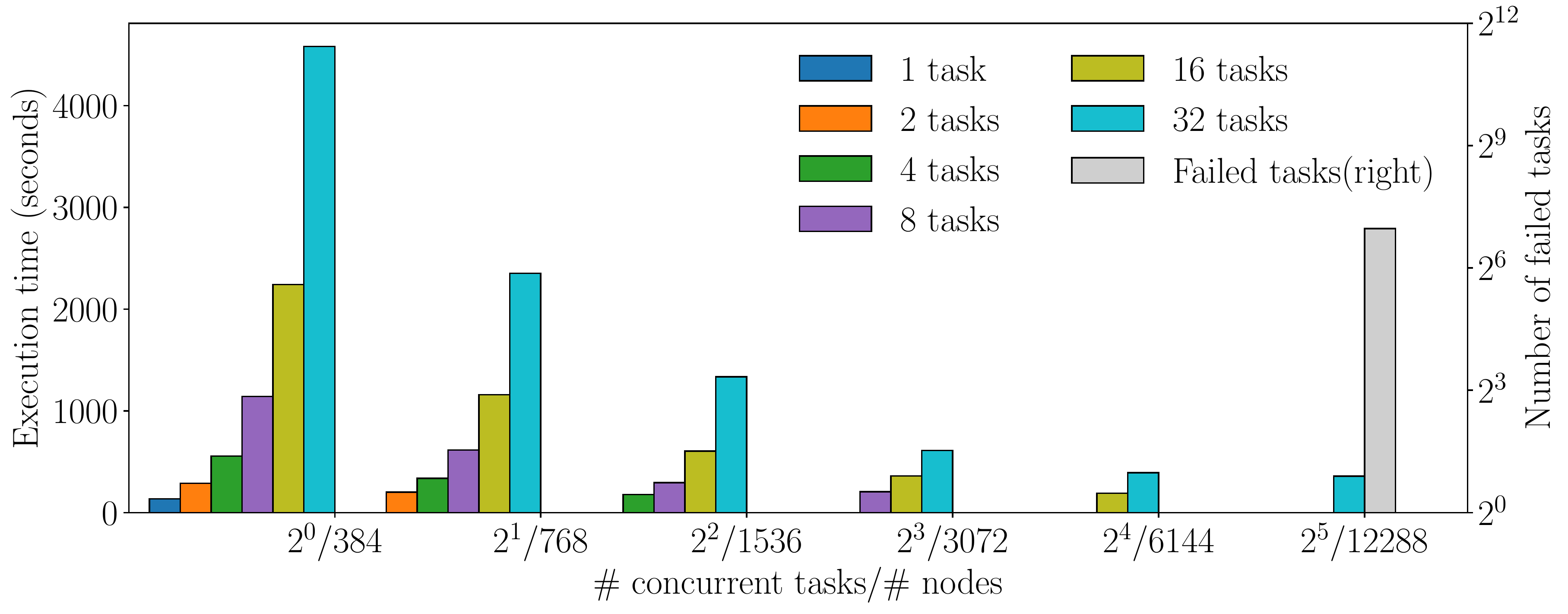}
  \caption{Task Execution Time of forward simulations using EnTK at
  various values of concurrency. 
  }\label{fig:specfem_concurrency}
\end{figure}

Fig.~\ref{fig:specfem_concurrency} shows that increasing concurrency leads to
a linear reduction of Task Execution Time, with a minimum of \(\approx\)180
seconds. Interestingly, reducing concurrency eliminates failures: we
encountered no failures in executions with up to \(2^4\) concurrent tasks and
6,144 nodes. At \(2^5\) concurrent tasks and 12,288 nodes, 50\% of the tasks
failed due to runtime issues.

EnTK automatically resubmitted failed tasks until they were successfully
executed. In the run with \(2^5\) tasks, EnTK attempted to run a total of 157
tasks. The resulting Task Execution Time was \(\approx\)360 seconds, similar
to that of a run with with \(2^4\) concurrent tasks
(Fig.~\ref{fig:specfem_concurrency}).

EnTK and RP enable reasoning and benchmarking the concurrency of an execution
without any change in the executable code. This gives insight on how to
tailor a given computational campaign on a specific CI\@. The insight gained
via our experiments can be immediately used in production: On Titan, forward
simulations are best executed with \(2^4\) concurrent tasks. Further,
fault-tolerance has an immediate impact on production runs, eliminating one
of the most limiting factor of the previous implementation of the workflow.

\subsubsection{Meteorological Probabilistic Forecasts}

We use EnTK to implement the \gls{AUA} algorithm to iteratively and
dynamically identify locations of the analogs. We also implement the
\textit{status quo} method of generating these analogs, i.e., random
selection of locations in each iteration. We perform experiments to compare
the two implementations and observe the speedup of the proposed algorithm. We
repeat the experiment 30 times for statistical accuracy, initializing both
implementations using the same initial random locations.

\begin{figure}[t]
    \centering
    \includegraphics[width=\columnwidth]{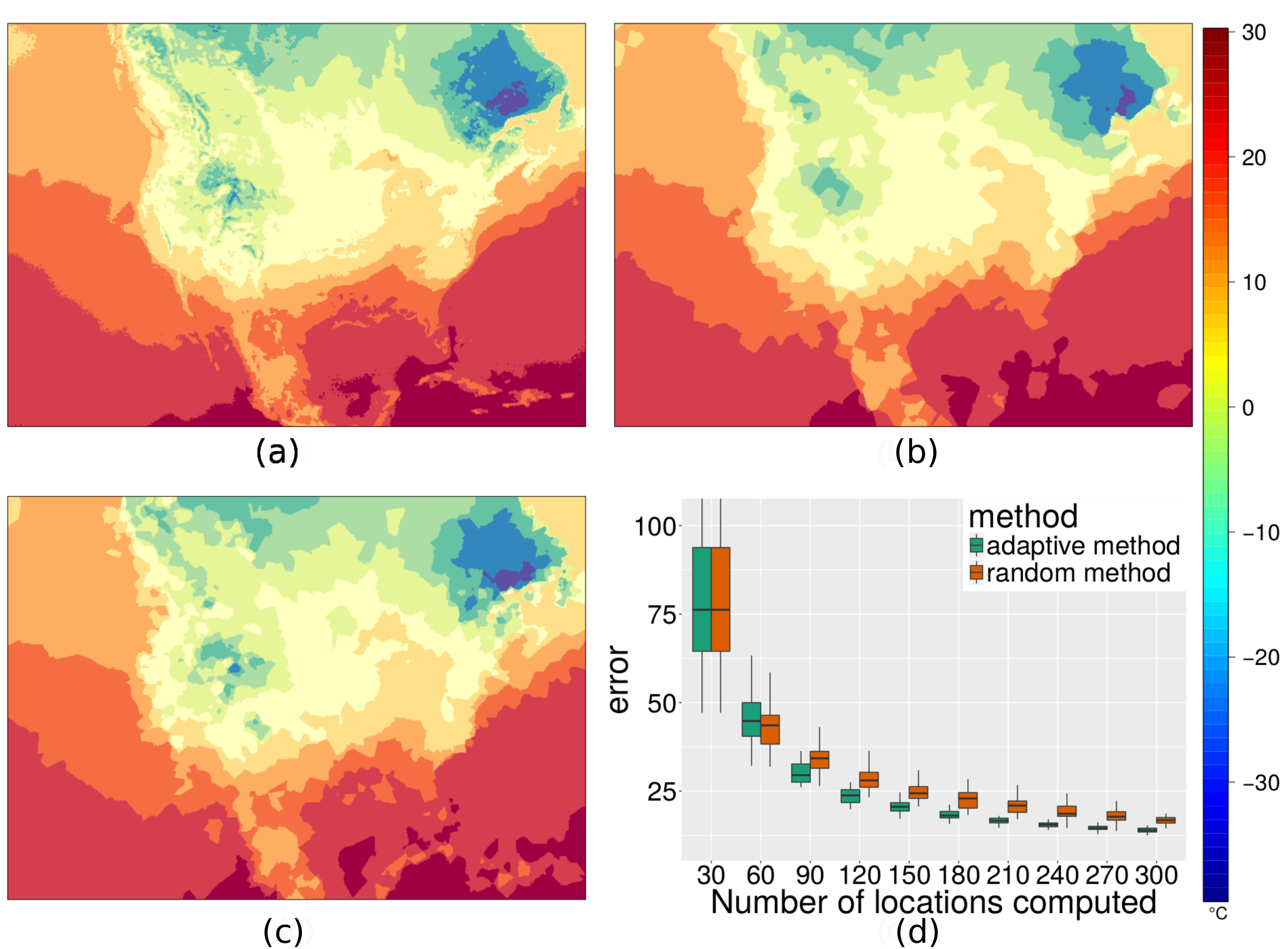}
    \caption{Predictions from random and adaptive methods. (a) theoretical
    true value, (b) the interpolated map from 1,800 randomly picked
    locations, (c) the interpolated map from 1,800 locations identified using
    AUA, (d) box plots of the errors for both
    implementations.}\label{fig:psu-error-plot}
\end{figure}

Fig.~\ref{fig:psu-error-plot} shows the prediction maps and errors obtained
from the two implementations. With 1,800 locations calculated for both
prediction maps (Fig.~\ref{fig:psu-error-plot}(b),
Fig.~\ref{fig:psu-error-plot}(c)), the \gls{AUA} algorithm generates a map
with certain areas that have a better representation of the analysis than the
map generated by a random selection of pixels.

The box plot in Fig.~\ref{fig:psu-error-plot}(d) shows the distribution of
the errors for the two implementations. The error converges faster in the
\gls{AUA} algorithm than in the random selection. The total amount of
potential locations (pixels) is 262,972; thus both implementations use a
small fraction of the available locations but the \gls{AUA} algorithm is
automatically steering the computation at each iteration. EnTK and RP avoid
the usual shortcoming of this approach: The evaluation required by the
steering can be implemented as a task and iterations do not wait in the HPC
queue, even if their number is unknown before execution. These results
suggest that the \gls{AUA} algorithm over random selection of points is well
suited for very large domains.

\section{Related Work}\label{sec:related_work}

Executing ensemble applications on HPC systems requires knowledge of
resource, data and execution management, specific to the HPC system. Several
``middleware''~\cite{stonebraker2002too} frameworks have been developed to
abstract execution details and enable execution of ensemble applications.
Software development kits such as gSOAP~\cite{aloisio2003secure} enable web
services for HPC applications. Ninf-G~\cite{tanaka2003ninf} and
OmniRPC~\cite{sato2003omnirpc} provide client/server-based frameworks for
distributed programming. These solutions provide methods to launch
application tasks on remote machines but leave the details of task
scheduling, resource and data management, and fault tolerance to the
user.


Hadoop and its ecosystem have been ported to HPC
systems~\cite{krishnan2011myhadoop,moody2013jummp,magpie}, enabling the use
of the MapReduce programming model. While some ensemble applications are
data-flow oriented and thus amenable to be implemented with MapReduce, EnTK
adopts a more flexible and coarse-grained notion of tasks, where a task in
EnTK can support multiple programming models, including MPI\@. Further, EnTK
does not assume a specific runtime system and, in conjunction with RP, can
use Hadoop on HPC~\cite{luckow2016hadoop}.


Feature-rich workflow systems such as Kepler~\cite{Ludascher2006},
Swift~\cite{wilde2011swift}, and Pegasus~\cite{Deelman2005} provide
end-to-end capabilities such as resource and execution management, fault
tolerance, monitoring and provenance. Encoding applications using these
systems requires acquiring specific knowledge, including learning new
languages and paradigms. Adapting these systems to user requirements is
non-trivial due to their feature richness and end-to-end design.
Ruffus~\cite{goodstadt2010ruffus}, COSMOS~\cite{gafni2014cosmos}, and GXP
Make~\cite{taura2013design} limit the capabilities and prioritize interface
simplicity. Galaxy~\cite{goecks2010galaxy}, Taverna~\cite{oinn2004taverna},
BioPipe~\cite{hoon2003biopipe}, and Copernicus~\cite{pronk2015molecular}
focus on providing tailored interfaces to domain scientists.

EnTK contributes (i) programmability, (ii) portability across CIs and RTSs,
and (iii) generality to the research on ensemble applications. These
applications can be expressed as workflows but their distinguishing patterns
permit a simplification of the graph structure while requiring better
handling of task parallelism and runtime adaptivity~\cite{entk}.
Consequently, EnTK exposes an API tailored towards encoding of ensemble
applications, focusing on task concurrency and sequentiality.

One of the limitations observed in the existing frameworks is that functional
and performance enhancements are localized to one framework and cannot be
easily ported to other systems. EnTK avoids framework lock-in by enabling
composability with diverse runtime systems and rapid development of
user-facing, special-purpose application libraries. In this way, EnTK builds
upon the idea of composing applications from execution patterns, also
explored by systems like Tigres, and extends it to middleware for
HPC for better programmability, portability, and generality.

\section{Conclusion}\label{sec:conclusion}
The results of our experiments show that the design and implementation of EnTK
meet the requirements of diverse use cases. The performance of EnTK is shown
to be invariant of workload and platform.  EnTK was shown to have
ideal weak and strong scaling up to currently required scales. Importantly,
any deviation from ideal scaling was explicable, and the causes are candidates
for future enhancements. The use of EnTK with Specfem at large scales on Titan
at ORNL led to unprecedented reductions in time-to-completion, insulation
against failures (e.g., hardware and software), and improved reliability.


Abstractions exposed by EnTK permit algorithmic innovations. For the
meteorological probabilistic forecast use case, the independence from direct
resource management permits new adaptive formulations of the Analog Ensemble
method, which in turn leads to improved accuracy in predictions, with reduced
time to completion and usage of compute resources.


We provide initial demonstrations of how EnTK has facilitated the full
potential of ensemble methods (``power of many''). EnTK will allow similar
methodological advances for other ensemble applications, which have so far
been hindered by the lack of suitable tools. EnTK is also a validation of the
building block approach to middleware: it is demonstrably extensible to
application specific frameworks in the upward direction~\cite{bb1}, as well
as being agnostic to the specific RTS below.

Having provided fundamental advances for ensemble applications at the largest
scales currently available ($\approx$66\% of Titans' nodes), EnTK will be
engineered to provide a pathway to pre-exascale levels without disruption in
production capabilities for users of Titan. Specifically, EnTK will provide
capabilities for: (i) dynamic mapping of tasks onto heterogeneous resources,
and (ii) and adaptive execution strategies to enable optimal resource
utilization.

\small{Acknowledgments: This work is supported by NSF ICER 1639694, DOE ASCR
DE-SC0016280 and an INCITE award to J.T. Resources from the Oak Ridge
Leadership Computing Facility at the Oak Ridge National Laboratory are
supported by the Office of Science of the U.S. Department of Energy, Contract
No. DE-AC05-00OR22725.}

\begin{spacing}{0.90}
\bibliographystyle{IEEEtran}
\bibliography{hpc-workflows-paper-y1}
\end{spacing}
\end{document}